\documentclass[12pt,a4paper]{article}
\usepackage{graphicx}
\begin{document}
\textwidth=135mm
 \textheight=200mm
\begin{center}
{\bfseries DVCS at HERMES}
\vskip 5mm
M. Murray$^{\dag}$ on behalf of the H{\sc ermes} Collaboration
\vskip 5mm
{\small {\it $^\dag$ Department of Physics and Astronomy, Univeristy
    of Glasgow, Glasgow, United Kingdom}} \\
\end{center}
\vskip 5mm
\centerline{\bf Abstract}
This talk explores the impact that the H{\sc ermes} experiment has had regarding knowledge of the Deeply Virtual Compton Scattering process. We discuss the various measurements that H{\sc ermes} has contributed to the library of DVCS knowledge [1--3], with focus in particular on the recent high-precision beam spin and charge asymmetries [4]. 
\vskip 10mm
\section{\label{sec:intro}Introduction}
The H{\sc ermes} experiment was a fixed-target experiment on the H{\sc
  era} ring in Hamburg, Germany. The experiment ran from 1995-2007 and
used the $27.6\,\textrm{GeV}$ electron/positron beam made available by
the H{\sc era} accelerator. A variety of polarised targets were used; the targets
under consideration in these proceedings were transversely polarised
(w.r.t. the beam direction) and polarisation averaged proton
targets. In this work we discuss some of the DVCS measurements taken by H{\sc ermes}.
\section{Exclusive Physics}

\begin{figure}
\centerline{  \includegraphics[width=1.2\textwidth]{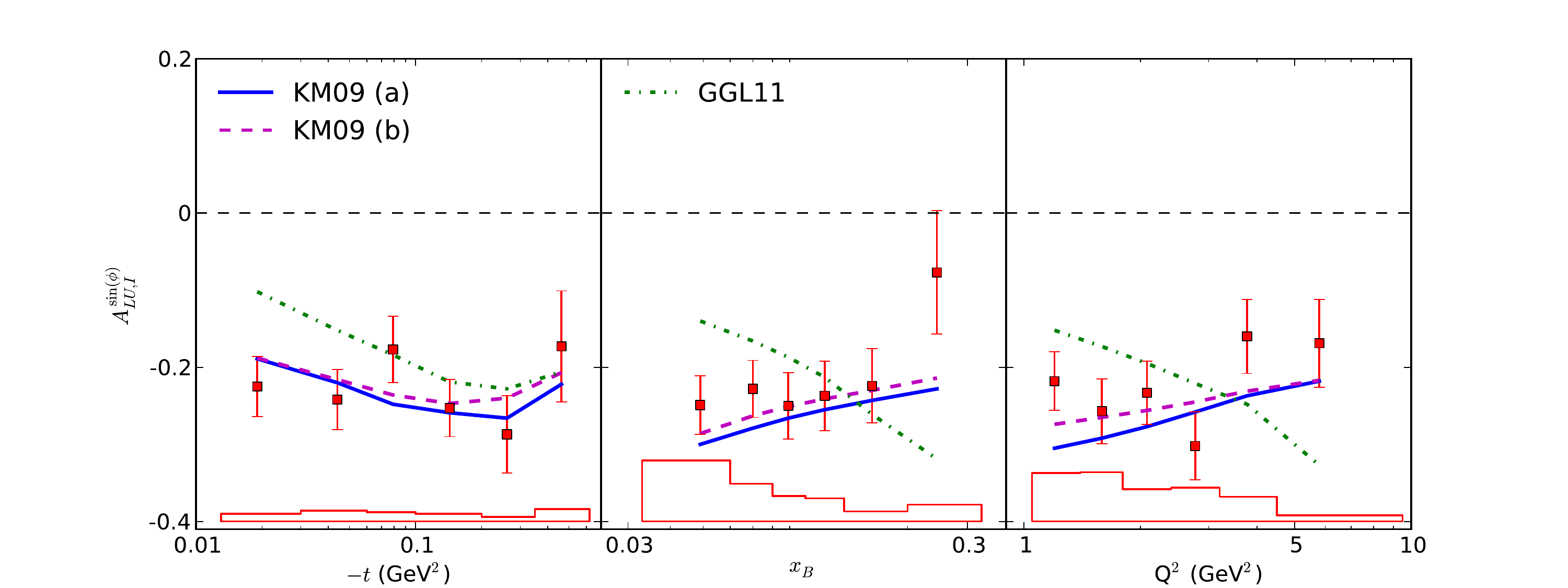}}
  \caption{The interference contribution to the first
    harmonic of the beam helicity asymmetry projected in six bins
    along the three kinematic dimensions $-t$, $x_{\textrm{B}}$ and
    $Q^2$. This measurement can be used to constrain  a part of the
    GPD $H$, the generalised parton distribution that reduces to the
    parton distribution $q(x)$ in its forward limit. Error bars are
    the statistical uncertainties; error bands are the systematic
    uncertainties. The curves on the figure come from \cite{km} and \cite{ggl}.}
  \label{fig:bsa}
\end{figure}

\begin{figure}
\centerline{  \includegraphics[width=1.2\textwidth]{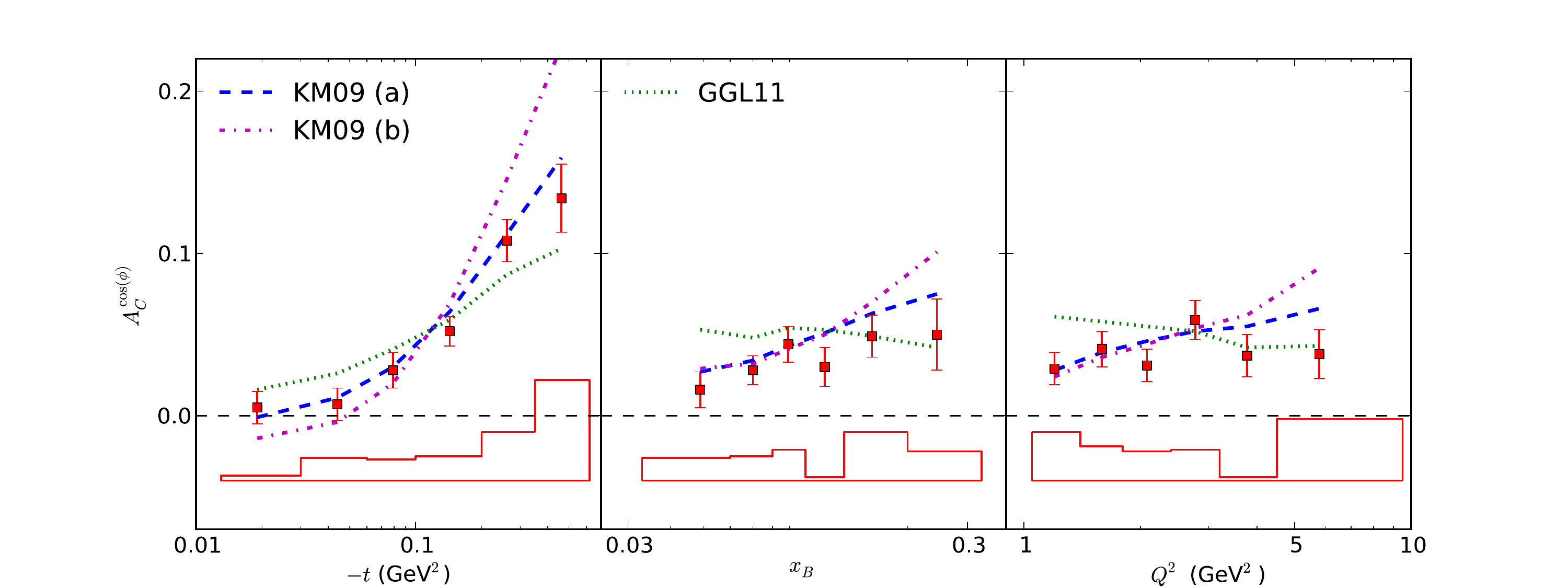}}
  \caption{The DVCS and interference contributions to the first
    harmonic of the beam charge asymmetry. This measurement can be
    used to constrain a part of the GPD $H$, the generalised parton distribution
    that reduces to the parton distribution $q(x)$ in its forward
    limit. The curves on the figure come from \cite{km} and \cite{ggl}.}
  \label{fig:bca}
\end{figure}

Knowledge of nucleon structure can be expanded by considering generalised
parton distributions (GPDs); access to GPDs can be achieved by measuring
particle production in scattering processes where the target nucleon
remains intact. The simplest process to understand is Deeply Virtual
Compton Scattering (DVCS), where a photon is produced by a parton from
the nucleon~\cite{hermesBSABCA1,hermesBSABCA2,hermesALT,hermesLTSA,hermesTTSA}. It is also possible to interpret measurements of produced
mesons in the GPD framework, e.g.~\cite{hermesMESON1,hermesMESON2,hermesMESON3,hermesMESON4}. The H{\sc ermes} experiment
recently produced DVCS measurements from the entire 1995-2007 data
set, where beam helicity and charge azimuthal asymmetries in the produced
particle distributions are measured, see figures~\ref{fig:bsa}
and~\ref{fig:bca} respectively. In these figures the measurements are
compared to values from a lightcone wavefront partonic model~\cite{km}
and from a flexible bag model~\cite{ggl}. Both of
these models receive input from PDF measurements and form factors. The
former is also fit to a large subset of the available world data,
whereas the latter's skewness dependence is determined solely from
measurements made at Halls A and B at Jefferson Lab.
The H{\sc ermes} experiment can distinguish the pure DVCS contribution
to the asymmetries from the contribution due to the interference with
the competing Bethe-Heitler process due to the presence of both beam
charges in the data set. These measurements are taken with a
missing-mass selection technique, where the target nucleon scatters
outside the H{\sc ermes} geometric acceptance and must be
reconstructed from measurements of the produced photon and scattered
lepton. This event selection technique allows a certain amount of
background into the data sample, mostly from events involving a
resonant state of the target nucleon. The collaboration has also
published measurements from a kinematically completely recostructed
data sample~\cite{hermesRECOIL} that was taken with additional
experimental equipment --- these measurements are discussed in other
contributions to this conference. Both figs.~\ref{fig:bsa}
and~\ref{fig:bca} show strong twist-2 amplitudes; in particular
fig.~\ref{fig:bsa} implies a steep drop-off of
$A^{\sin(phi)}_{LU,I}$ at small values of $t$ as the amplitude must
become zero-valued at the point $t=0\,\textrm{GeV}^2$ as a matter of definition.

\subsection*{Acknowledgements}
 We gratefully acknowledge the D{\sc esy} management for its support, the staff
at D{\sc esy} and the collaborating institutions for their significant effort,
and our national funding agencies and the EU FP7 (HadronPhysics2, Grant
Agreement number 227431) for financial support.

\end{document}